# Bayes' Bluff: Opponent Modelling in Poker

**Finnegan Southey, Michael Bowling, Bryce Larson, Carmelo Piccione, Neil Burch, Darse Billings, Chris Rayner**
Department of Computing Science
University of Alberta
Edmonton, Alberta, Canada T6G 2E8
{finnegan,bowling,larson,carm,burch,darse,rayner}@cs.ualberta.ca

## Abstract

Poker is a challenging problem for artificial intelligence, with non-deterministic dynamics, partial observability, and the added difficulty of unknown adversaries. Modelling all of the uncertainties in this domain is not an easy task. In this paper we present a Bayesian probabilistic model for a broad class of poker games, separating the uncertainty in the game dynamics from the uncertainty of the opponent's strategy. We then describe approaches to two key subproblems: (i) inferring a posterior over opponent strategies given a prior distribution and observations of their play, and (ii) playing an appropriate response to that distribution. We demonstrate the overall approach on a reduced version of poker using Dirichlet priors and then on the full game of Texas hold'em using a more informed prior. We demonstrate methods for playing effective responses to the opponent, based on the posterior.

## 1 Introduction

The game of poker presents a serious challenge to artificial intelligence research. Uncertainty in the game stems from partial information, unknown opponents, and game dynamics dictated by a shuffled deck. Add to this the large space of possible game situations in real poker games such as Texas hold'em, and the problem becomes very difficult indeed. Among the more successful approaches to playing poker is the game theoretic approach, approximating a Nash equilibrium of the game via linear programming [5, 1]. Even when such approximations are good, Nash solutions represent a pessimistic viewpoint in which we face an optimal opponent. Human players, and even the best computer players, are certainly not optimal, having idiosyncratic weaknesses that can be exploited to obtain higher payoffs than the Nash value of the game. Opponent modelling attempts to capture these weaknesses so they can be exploited in subsequent play.

Existing approaches to opponent modelling have employed a variety of approaches including reinforcement learning [4], neural networks [2], and frequentist statistics [3]. Additionally, earlier work on using Bayesian models for poker [6] attempted to classify the opponent's hand into one of a variety of broad hand classes. They did not model uncertainty in the opponent's strategy, using instead an explicit strategy representation. The strategy was updated based on empirical frequencies of play, but they reported little improvement due to this updating. We present a general Bayesian probabilistic model for *hold 'em* poker games, completely modelling the uncertainty in the game and the opponent.

We start by describing hold'em style poker games in general terms, and then give detailed descriptions of the casino game Texas hold'em along with a simplified research game called Leduc hold'em for which game theoretic results are known. We formally define our probabilistic model and show how the posterior over opponent strategies can be computed from observations of play. Using this posterior to exploit the opponent is non-trivial and we discuss three different approaches for computing a response. We have implemented the posterior and response computations in both Texas and Leduc hold'em, using two different classes of priors: independent Dirichlet and an informed prior provided by an expert. We show results on the performance of these Bayesian methods, demonstrating that they are capable of quickly learning enough to exploit an opponent.

## 2 Poker

There are many variants of poker.[1] We will focus on *hold 'em*, particularly the heads-up limit game (i.e., two players with pre-specified bet and raise amounts). A single hand consists of a number of rounds. In the first round, players are dealt a fixed number of private cards. In all rounds, some fixed number (possibly zero) of shared, public *board* cards are revealed. The dealing and/or revealing of cards is followed by betting. The betting involves alternating decisions, where each player can either *fold* (f), *call* (c), or *raise* (r). If a player folds, the hand ends and the other player wins the *pot*. If a player calls, they place into the pot an amount to match what the other player has already placed in the pot (possibly nothing). If a player raises, they match the other player's total and then put in an additional fixed amount. The players alternate until a player folds, ending the hand, or a player calls (as long as the call is not the first action of the round), continuing the hand to the next round.

[1] A more thorough introduction of the rules of poker can be found in [2].

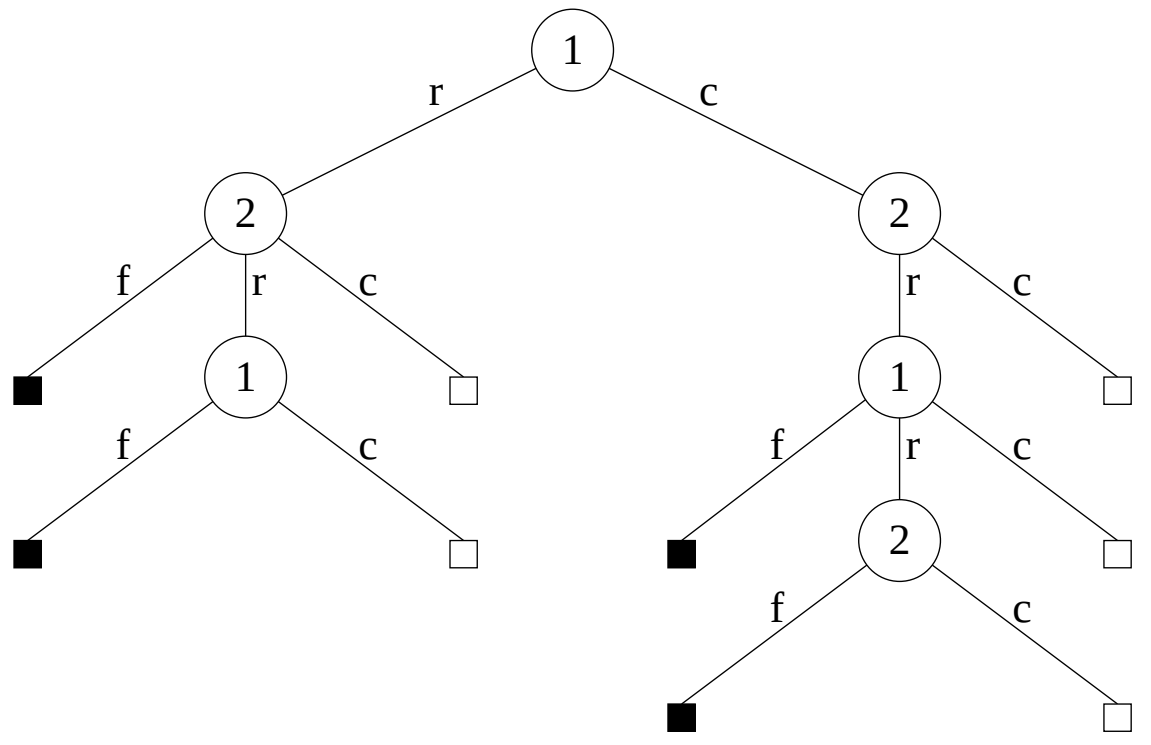

Figure 1: An example decision tree for a single *betting round* in poker with a two-bet maximum. Leaf nodes with open boxes continue to the next round, while closed boxes end the hand.

There is a limit on the number of raises (or bets) per round, so the betting sequence has a finite length. An example decision tree for a single round of betting with a two-bet maximum is shown in Figure 1. Since folding when both players have equal money in the pot is dominated by the call action, we do not include this action in the tree. If neither player folds before the final betting round is over, a *showdown* occurs. The players reveal their private cards and the player who can make the strongest poker hand with a combination of their private cards and the public board cards wins the pot.

Many games can be constructed with this simple format for both analysis (*e.g.*, Kuhn poker [7] and Rhode Island hold'em [9]) and human play. We focus on the commonly played variant, Texas hold 'em, along with a simplified and more tractable game we constructed called Leduc hold 'em.

**Texas Hold 'Em.** The most common format for hold 'em is "Texas Hold'em", which is used to determine the human world champion and is widely considered the most strategically complex variant. A standard 52-card deck is used. There are four betting rounds. In the first round, the players are dealt two private cards. In the second round (or *flop*), three board cards are revealed. In the third round (*turn*) and fourth round (*river*), a single board card is revealed. We use a four-bet maximum, with fixed raise amounts of 10 units in the first two rounds and 20 units in the final two rounds. Finally, *blind bets* are used to start the first round. The first player begins the hand with 5 units in the pot and the second player with 10 units.

**Leduc Hold 'Em.** We have also constructed a smaller version of hold 'em, which seeks to retain the strategic elements of the large game while keeping the size of the game tractable. In Leduc hold 'em, the deck consists of two suits with three cards in each suit. There are two rounds. In the first round a single private card is dealt to each player. In the second round a single board card is revealed. There is a two-bet maximum, with raise amounts of 2 and 4 in the first and second round, respectively. Both players start the first round with 1 already in the pot.

**Challenges.** The challenges introduced by poker are many. The game involves a number of forms of uncertainty, including stochastic dynamics from a shuffled deck, imperfect information due to the opponent's private cards, and, finally, an unknown opponent. These uncertainties are individually difficult and together the difficulties only escalate. A related challenge is the problem of folded hands, which amount to partial observations of the opponent's decision-making contexts. This has created serious problems for some opponent modelling approaches and our Bayesian approach will shed some light on the additional challenge that fold data imposes. A third key challenge is the high variance of payoffs, also known as *luck*. This makes it difficult for a program to even assess its performance over short periods of time. To aggravate this difficulty, play against human opponents is necessarily limited. If no more than two or three hundred hands are to be played in total, opponent modelling must be effective using only very small amounts of data. Finally, Texas hold'em is a very large game. It has on the order of $10^{18}$ states [1], which makes even straightforward calculations, such as best response, non-trivial.

## 3 Modelling the Opponent

We will now describe our probabilistic model for poker. In all of the following discussion, we will assume that Player 1 (P1) is modelling its opponent, Player 2 (P2), and that all incomplete observations due to folding are from P1's perspective.

### 3.1 Strategies

In game theoretic terms, a player makes decisions at *information sets*. In poker, information sets consist of the actions taken by all players so far, the public cards revealed so far, and the player's own private cards. A *behaviour strategy* specifies a distribution over the possible actions for every information set of that player. Leaving aside the precise form of these distributions for now, we denote P1's complete strategy by $\alpha$ and P2's by $\beta$.

We make the following simplifying assumptions regarding the player strategies. First, P2's strategy is stationary. This is an unrealistic assumption but modelling stationary opponents in full-scale poker is still an open problem. Even the most successful approaches make the same assumption or use simple methods such as decaying histories to accommodate opponent drift. However, we believe this framework can be naturally extended to dynamic opponents by constructing priors that explicitly model changes in opponent strategy. The second assumption is that the players' strategies are independent. More formally, $P(\alpha, \beta) = P(\alpha)P(\beta)$. This assumption, implied by the stationarity, is also unrealistic. Hower, modelling opponents that learn, and effectively deceiving them, is a difficult task even in very small games and we defer such efforts until we are sure of effective stationary opponent modelling. Finally, we assume the deck is uniformly distributed, *i.e.*, the game is fair. These assumptions imply that all hands are i.i.d. given the strategies of the players.

### 3.2 Hands

The following notation is used for hand information. We consider a hand, $H$, with $k$ decisions by each player. Each hand, as observed by an oracle with perfect information, is a tuple $H = (C, D, R_{1:k}, A_{1:k}, B_{1:k})$ where,

- $C$ and $D$ denote P1 and P2's private cards,
- $R_i$ is the set (possibly empty) of public cards dealt before either player makes their $i$th decision, and
- $A_i$ and $B_i$ denote P1 and P2's $i$th decisions (fold, call or raise).

We can model any limit hold'em style poker with these variables. A hand runs to at most $k$ decisions. The fact that particular hands may have fewer real decisions (*e.g.*, a player may call and end the current betting round, or fold and end the hand) can be handled by padding the decisions with specific values (*e.g.*, once a player has folded all subsequent decisions by both players are assumed to be folds). Probabilities in the players' strategies for these padding decisions are forced to 1. Furthermore, the public cards for a decision point ($R_i$) can be the empty set, so that multiple decisions constituting a single betting round can occur between revealed public cards. These special cases are quite straightforward and allow us to model the variable length hands found in real games with fixed length tuples.

### 3.3 Probability of Observations

Suppose a hand is fully observed, *i.e.*, a showdown occurs. The probability of a particular showdown hand $H_s$ occurring given the opponent's strategy is, [2]

$$
\begin{aligned}
&P(H_s|\beta) \\
&= P(C, D, R_{1:k}, A_{1:k}, B_{1:k}|\beta) \\
&= P(D|C)P(C) \prod_{i=1}^{k} \big[ P(B_i|D, R_{1:i}, A_{1:i}, B_{1:i-1}, \beta) \\
&\qquad P(A_i|C, R_{1:i}, A_{1:i-1}, B_{1:i-1}) \\
&\qquad P(R_i|C, D, R_{1:i-1}) \big] \\
&= P(D|C)P(C) \prod_{i=1}^{k} \big[ \alpha_{Y_i,C,A_i} \; \beta_{Z_i,D,B_i} \\
&\qquad P(R_i|C, D, R_{1:i-1}) \big] \\
&= p_{\text{showcards}} \prod_{i=1}^{k} \alpha_{Y_i,C,A_i} \; \beta_{Z_i,D,B_i} \\
&\propto \prod_{i=1}^{k} \beta_{Z_i,D,B_i},
\end{aligned}
$$

where for notational convenience, we separate the information sets for P1 (P2) into its public part $Y_i$ ($Z_i$) and its private part $C$ ($D$). So,

$$
\begin{aligned}
Y_i &= (R_{1:i}, A_{1:i-1}, B_{1:i-1}) \\
Z_i &= (R_{1:i}, A_{1:i}, B_{1:i-1}).
\end{aligned}
$$

In addition, $\alpha_{Y_i,C,A_i}$ is the probability of taking action $A_i$ in the information set $(Y_i, C)$, dictated by P1's strategy, $\alpha$. A similar interpretation applies to the subscripted $\beta$. $p_{\text{showcards}}$ is a constant that depends only on the number of cards dealt to players and the number of public cards revealed. This simplification is possible because the deck has uniform distribution and the number of cards revealed is the same for all showdowns. Notice that the final unnormalized probability depends only on $\beta$.

Now consider a hand where either player folds. In this case, we do not observe P2's private cards, $D$. We must marginalize away this hidden variable by summing over all possible sets of cards P2 could hold.

[2] Strictly speaking, this should be $P(H|\alpha, \beta)$ but we drop the conditioning on $\alpha$ here and elsewhere to simplify the notation.

The probability of a particular fold hand $H_\text{f}$ occurring is,

$$
\begin{aligned}
&P(H_\text{f}|\beta) \\
&= P(C, R_{1:k}, A_{1:k}, B_{1:k}|\beta) \\
&= P(C)\sum_{D} P(D|C) \prod_{i=1}^{k} \big[\, P(B_i|D, R_{1:i}, A_{1:i}, B_{1:i-1}, \beta) \\
&\qquad\qquad\qquad\qquad P(A_i|C, R_{1:i}, A_{1:i-1}, B_{1:i-1}) \\
&\qquad\qquad\qquad\qquad P(R_i|C, D, R_{1:i-1}) \,\big] \\
&= p_\text{foldcards}(H_\text{f}) \left[\prod_{i=1}^{k} \alpha_{Y_i,C,A_i}\right] \sum_{D'} \prod_{i=1}^{k} \beta_{Z_i,D',B_i} \\
&\propto \sum_{D'} \prod_{i=1}^{k} \beta_{Z_i,D',B_i}
\end{aligned}
$$

where $D'$ are sets of cards that P2 could hold given the observed $C$ and $R$ (*i.e.*, all sets $D$ that do not intersect with $C \cup R$), and $p_\text{foldcards}(H_\text{f})$ is a function that depends only on the number of cards dealt to the players and the number of public cards revealed before the hand ended. It does not depend on the specific cards dealt or the players' strategies. Again, the unnormalized probability depends only on $\beta$.

### 3.4 Posterior Distribution Over Opponent Strategies

Given a set $\mathcal{O} = \mathcal{O}_\text{s} \cup \mathcal{O}_\text{f}$ of observations, where $\mathcal{O}_\text{s}$ are the observations of hands that led to showdowns and $\mathcal{O}_\text{f}$ are the observations of hands that led to folds, we wish to compute the posterior distribution over the space of opponent strategies. A simple application of Bayes' rule gives us,

$$
\begin{aligned}
P(\beta|\mathcal{O}) &= \frac{P(\mathcal{O}|\beta)P(\beta)}{P(\mathcal{O})} \\
&= \frac{P(\beta)}{P(\mathcal{O})} \prod_{H_\text{s}\in\mathcal{O}_\text{s}} P(H_\text{s}|\beta) \prod_{H_\text{f}\in\mathcal{O}_\text{f}} P(H_\text{f}|\beta) \\
&\propto P(\beta) \prod_{H_\text{s}\in\mathcal{O}_\text{s}} P(H_\text{s}|\beta) \prod_{H_\text{f}\in\mathcal{O}_\text{f}} P(H_\text{f}|\beta)
\end{aligned}
$$

## 4 Responding to the Opponent

Given a posterior distribution over the opponent's strategy space, the question of how to compute an appropriate response remains. We present several options with varying computational burdens. In all cases we compute a response at the beginning of the hand and play it for the entire hand.

### 4.1 Bayesian Best Response

The fully Bayesian answer to this question is to compute the best response to the entire distribution. We will call this the *Bayesian Best Response* (BBR). The objective here is to maximize the expected value over all possible hands and opponent strategies, given our past observations of hands. We start with a simple objective,

$$
\begin{aligned}
\alpha^\text{BBR} &= \operatorname*{argmax}_{\alpha} E_{H|\mathcal{O}} V(H) \\
&= \operatorname*{argmax}_{\alpha} \sum_{H\in\mathcal{H}} V(H) P(H|\mathcal{O}, \alpha) \\
&= \operatorname*{argmax}_{\alpha} \sum_{H\in\mathcal{H}} V(H) \int_{\beta} P(H|\alpha, \beta, \mathcal{O}) P(\beta|\mathcal{O}) \\
&= \operatorname*{argmax}_{\alpha} \sum_{H\in\mathcal{H}} V(H) \int_{\beta} P(H|\alpha, \beta, \mathcal{O}) P(\mathcal{O}|\beta) P(\beta) \\
&= \operatorname*{argmax}_{\alpha} \sum_{H\in\mathcal{H}} V(H) \int_{\beta} \big[\, P(H|\alpha, \beta, \mathcal{O}) P(\beta) \\
&\qquad \prod_{H_\text{s}\in\mathcal{O}_\text{s}} \prod_{i=1}^{k} \beta_{Z_i,D,B_i} \\
&\qquad \prod_{H_\text{f}\in\mathcal{O}_\text{f}} \sum_{D'} \prod_{i=1}^{k} \beta_{Z_i,D,B_i} \big]
\end{aligned}
$$

where $\mathcal{H}$ is the set of all possible perfectly observed hands (in effect, the set of all hands that could be played). Although not immediately obvious from the equation above, one algorithm for computing Bayesian best response is a form of Expectimax [8], which we will now describe.

Begin by constructing the tree of possible observations in the order they would be observed by P1, including P1's cards, public cards, P2's actions, and P1's actions. At the bottom of the tree will be an enumeration of P2's cards for both showdown and fold outcomes. We can backup values to the root of the tree while computing the best response strategy. For a leaf node the value should be the payoff to P1 multiplied by the probability of P2's actions reaching this leaf given the posterior distribution over strategies. For an internal node, calculate the value from its children based on the type of node. For a P2 action node or a public card node, the value is the sum of the children's values. For a P1 action node, the value is the maximum of its children's values, and the best-response strategy assigns probability one to the action that leads to the maximal child for that node's information set. Repeat until every node has been assigned a value, which implies that every P1 information set has been assigned an action. More formally Expectimax computes the following value for the root of the tree,

$$
\sum_{R_1} \max_{A_1} \sum_{B_1} \cdots \sum_{R_k} \max_{A_k} \sum_{B_k} \sum_{D} V(H) \int_{\beta} \prod_{i=1}^{k} \beta_{Z_i,D,B_i} P(\mathcal{O}|\beta) P(\beta)
$$

This corresponds to Expectimax, with the posterior inducing a probability distribution over actions at P2's action nodes.

It now remains to prove that this version of Expectimax computes the BBR. This will be done by showing that,

$$\max_{\alpha} \sum_{H \in \mathcal{H}} V(H) \int_{\beta} P(H|\alpha, \beta, \mathcal{O}) P(\mathcal{O}|\beta) P(\beta)$$
$$\leq \sum_{R_1} \max_{A_1} \sum_{B_1} \cdots \sum_{R_k} \max_{A_k} \sum_{B_k} \sum_{D} V(H)$$
$$\int_{\beta} \prod_{i=1}^{k} \beta_{Z_i, D, B_i} P(\mathcal{O}|\beta) P(\beta)$$

First we rewrite $\max_{\alpha} \sum_H$ as,

$$\max_{\alpha(1)} \cdots \max_{\alpha(k)} \sum_{R_1} \sum_{A_1} \sum_{B_1} \cdots \sum_{R_k} \sum_{A_k} \sum_{B_k} \sum_{D},$$

where $\max_{\alpha(i)}$ is a max over the set of all parameters in $\alpha$ that govern the $i$th decision. Then, because $\max_x \sum_y f(x, y) \leq \sum_y \max_x f(x, y)$, we get,

$$\max_{\alpha(1)} \cdots \max_{\alpha(k)} \sum_{R_1} \sum_{A_1} \sum_{B_1} \cdots \sum_{R_k} \sum_{A_k} \sum_{B_k} \sum_{D}$$
$$\leq \max_{\alpha(2)} \cdots \max_{\alpha(k)} \sum_{R_1} \max_{\alpha(1)} \sum_{A_1} \sum_{B_1} \cdots \sum_{R_k} \sum_{A_k} \sum_{B_k} \sum_{D}$$
$$\leq \sum_{R_1} \max_{\alpha(1)} \sum_{A_1} \sum_{B_1} \cdots \sum_{R_k} \max_{\alpha(k)} \sum_{A_k} \sum_{B_k} \sum_{D}$$

Second, we note that,

$$\int_{\beta} P(H|\alpha, \beta, \mathcal{O}) P(\mathcal{O}|\beta) P(\beta)$$
$$\propto \prod_{i=1}^{k} \alpha_{Y_i, C, A_i} \int_{\beta} \prod_{i=1}^{k} \beta_{Z_i, D, B_i}$$

We can distribute parameters from $\alpha$ to obtain,

$$\sum_{R_1} \max_{\alpha(1)} \sum_{A_1} \alpha_{Y_1, C, A_1} \sum_{B_1} \cdots$$
$$\sum_{R_k} \max_{\alpha(k)} \sum_{A_k} \alpha_{Y_k, C, A_k} \sum_{B_k} \sum_{D}$$
$$\int_{\beta} \prod_{i=1}^{k} \beta_{Z_i, D, B_i} P(\mathcal{O}|\beta) P(\beta)$$
$$= \sum_{R_1} \max_{A_1} \sum_{B_1} \cdots \sum_{R_k} \max_{A_k} \sum_{B_k} \sum_{D}$$
$$\int_{\beta} \prod_{i=1}^{k} \beta_{Z_i, D, B_i} P(\mathcal{O}|\beta) P(\beta),$$

which is the Expectimax algorithm. This last step is possible because parameters in $\alpha$ must sum to one over all possible actions at a given information set. The maximizing parameter setting is to take the highest-valued action with probability 1.

Computing the integral over opponent strategies depends on the form of the prior but is difficult in any event. For Dirichlet priors (see Section 5), it is possible to compute the posterior exactly but the calculation is expensive except for small games with relatively few observations. This makes the exact BBR an ideal goal rather than a practical approach. For real play, we must consider approximations to BBR.

One straightforward approach to approximating BBR is to approximate the integral over opponent strategies by importance sampling using the prior as the proposal distribution:

$$\int_{\beta} P(H|\alpha, \beta, \mathcal{O}) P(\mathcal{O}|\beta) P(\beta) \approx \sum_{\tilde{\beta}} P(H|\alpha, \tilde{\beta}, \mathcal{O}) P(\mathcal{O}|\tilde{\beta})$$

where the $\tilde{\beta}$ are sampled from the prior, $\tilde{\beta} \sim P(\beta)$. More effective Monte Carlo techniques might be possible, depending on the prior used.

Note that $P(\mathcal{O}|\tilde{\beta})$ need only be computed once for each $\tilde{\beta}$, while the much smaller computation $P(H|\alpha, \tilde{\beta}, \mathcal{O})$ must be computed for every possible hand. The running time of computing the posterior for a strategy sample scales linearly in the number of samples used in the approximation and the update is constant time for each hand played. This tractability facilitates other approximate response techniques.

### 4.2 Max A Posteriori Response

An alternate goal to BBR is to find the *max a posteriori* (MAP) strategy of the opponent and compute a best response to that strategy. Computing a true MAP strategy for the opponent is also hard, so it is more practical to approximate this approach by sampling a set of strategies from the prior and finding the most probable amongst that set. This sampled strategy is taken to be an estimate of a MAP strategy and a best response to it is computed and played. MAP is potentially dangerous for two reasons. First, if the distribution is multimodal, a best response to any single mode may be suboptimal. Second, repeatedly playing any single strategy may never fully explore the opponent's strategy.

### 4.3 Thompson's Response

A potentially more robust alternative to MAP is to sample a strategy from the posterior distribution and play a best response to that strategy. As with BBR and MAP, sampling the posterior directly may be difficult. Again we can use importance sampling, but in a slightly different way. We sample a set of opponent strategies from the prior, compute their posterior probabilities, and then sample one strategy according to those probabilities.

$$P(i) = \frac{P(\tilde{\beta}_i | H, \mathcal{O})}{\sum_j P(\tilde{\beta}_j | H, \mathcal{O})}$$

This was first proposed by Thompson [10]. Thompson's has some probability of playing a best-response to any non-zero probability opponent strategy and so offers more robust exploration.

## 5 Priors

As with all Bayesian approaches, the resulting performance and efficiency depends on the choice of prior. Obviously the prior should capture our beliefs about the strategy of our opponent. The form of the prior also determines the tractability of (i) computing the posterior, and (ii) responding with the model. As the two games of hold 'em are considerably different in size, we explore two different priors.

**Independent Dirichlet.** The game of Leduc hold 'em is sufficiently small that we can have a fully parameterized model, with well-defined priors at every information set. Dirichlet distributions offer a simple prior for multinomials, which is a natural description for action probabilities. Any strategy (in behavioural form) specifies a multinomial distribution over legal actions for every information set. Our prior over strategies, which we will refer to as an *independent Dirichlet prior*, consists of independent Dirichlet distributions for each information set. We are using Dirichlet$(2, 2, 2)$ distributions, whose mode is the multinomial $(1/3, 1/3, 1/3)$ over fold, call, and raise.

**Informed.** In the Texas hold 'em game, priors with independent distributions for each information set are both intractable and ineffective. The size of the game virtually guarantees that one will never see the same information set twice. Any useful inference must be across information sets and the prior must encode how the opponent's decisions at information sets are likely to be correlated. We therefore employ an expert defined prior that we will refer to as an *informed prior*.

The informed prior is based on a ten dimensional recursive model. That is, by specifying values for two sets of five intuitive parameters (one set for each player), a complete strategy is defined. Table 1 summarizes the expert defined meaning of these five parameters. From the modelling perspective, we can simply consider this expert abstraction to provide us with a mapping from some low-dimensional parameter space to the space of all strategies. By defining a density over this parameter space, the mapping specifies a resulting density over behaviour strategies, which serves as our prior. In this paper we use an independent Gaussian distribution over the parameter space with means and variances chosen by a domain expert. We omit further details of this model because it is not the intended contribution of this paper but rather a means to demonstrate our approach on the large game of Texas hold'em.

Table 1: The five parameter types in the informed prior parameter space. A corresponding set of five are required to specify the opponent's model of how we play.

| Parameter | Description |
|---|---|
| $r_0$ | Fraction of opponent's strength distribution that must be exceeded to raise after \$0 bets (*i.e.*, to initiate betting). |
| $r_1$ | Fraction of opponent's strength distribution that must be exceeded to raise after >\$0 bets (*i.e.*, to raise). |
| $b$ | Fraction of the game-theoretic optimal bluff frequency. |
| $f$ | Fraction of the game-theoretic optimal fold frequency. |
| $t$ | Trap or slow-play frequency. |

## 6 Experimental Setup

We tested our approach on both Leduc hold'em with the Dirichlet prior and Texas hold'em with the informed prior. For the Bayesian methods, we used all three responses (BBR, MAP, and Thompson's) on Leduc and the Thompson's response for Texas (BBR has not been implemented for Texas and MAP's behaviour is very similar to Thompson's, as we will describe below). For all Bayesian methods, 1000 strategies were sampled from the prior at the beginning of each trial and used throughout the trial.

We have several players for our study. *Opti* is a Nash (or minimax) strategy for the game. In the case of Leduc, this has been computed exactly. We also sampled opponents from our priors in both Leduc and Texas, which we will refer to as *Priors*. In the experiments shown, a new opponent was sampled for each trial (200 hands), so results are averaged over many samples from the priors. Both Priors and Opti are static players. Finally, for state-of-the-art opponent modelling, we used *Frequentist*, (also known as *Vexbot*) described fully in [3] and implemented for Leduc.

All experiments consisted of running two players against each other for two hundred hands per trial and recording the *bankroll* (accumulated winnings/losses) at each hand. These results were averaged over multiple trials (1000 trials for all Leduc experiments and 280 trials for the Texas experiments). We present two kinds of plots. The first is simply average bankroll per number of hands played. A straight line on such a plot indicates a constant winning rate. The second is the average winning rate per number of hands played (*i.e.*, the first derivative of the aver-age bankroll). This allows one to see the effects of learning more directly, since positive changes in slope indicate improved exploitation of the opponent. Note that winning rates for small numbers of hands are very noisy, so it is difficult to interpret the early results. All results are expressed in raw pot units (*e.g.*, bets in the first and second rounds of Leduc are 2 and 4 units respectively).

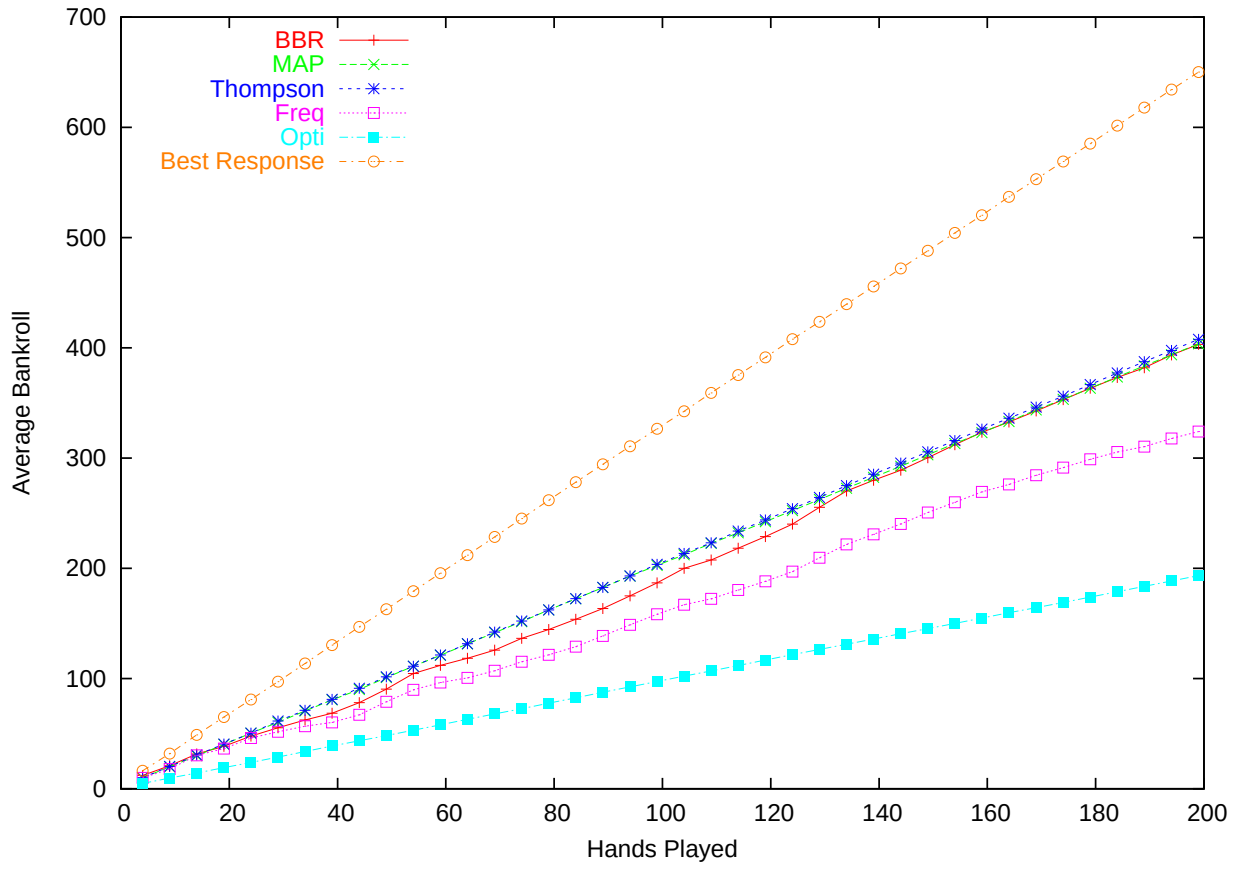

Figure 2: Leduc hold'em: Avg. Bankroll per hands played for BBR, MAP, Thompson's, Opti, and Frequentist vs. Priors.

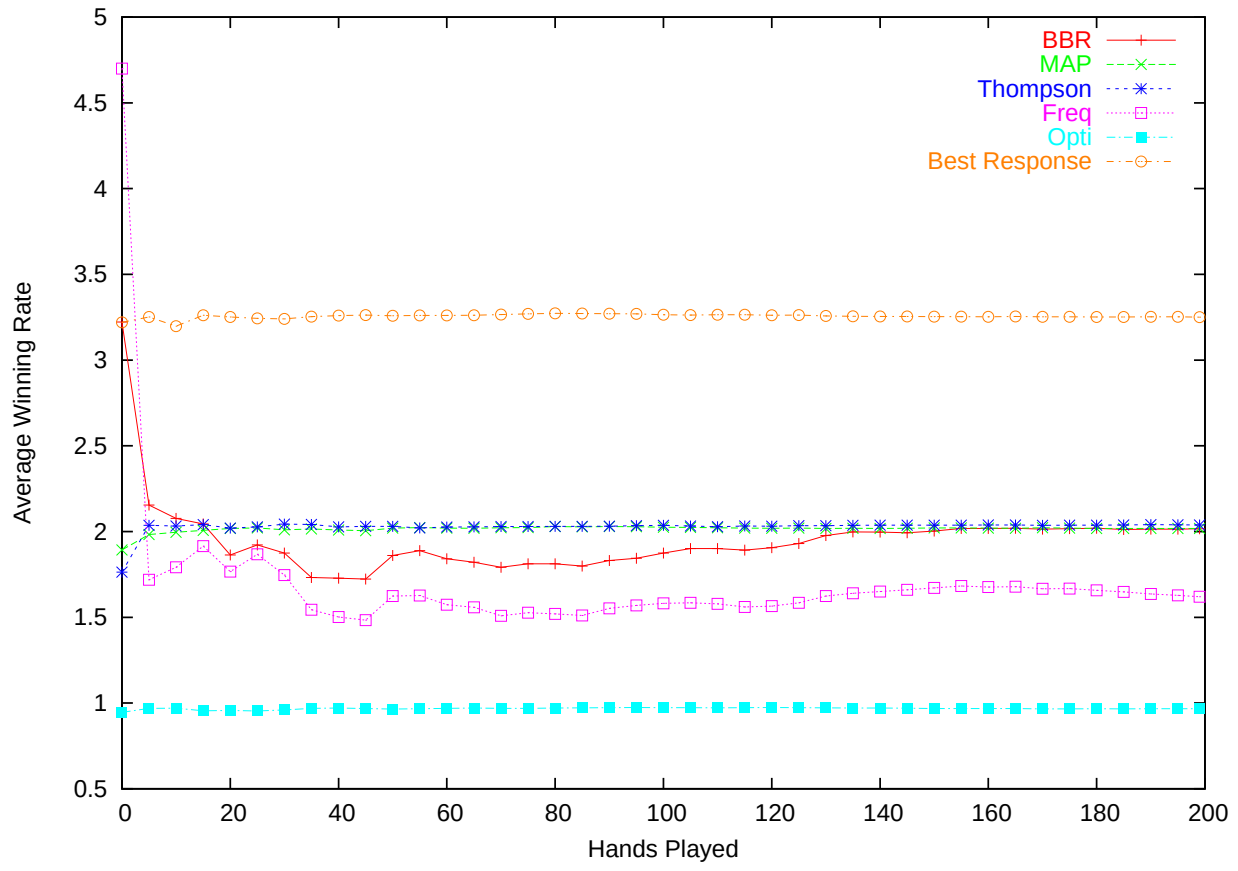

Figure 3: Leduc hold'em: Avg. Winning Rate per hands played for BBR, MAP, Thompson's, Opti, and Frequentist vs. Priors.

# 7 Results

## 7.1 Leduc Hold'em

Figures 2 and 3 show the average bankroll and average winning rate for Leduc against opponents sampled from the prior (a new opponent each trial). For such an opponent, we can compute a best response, which represents the best possible exploitation of the opponent. In complement, the Opti strategy shows the most conservative play by assuming that the opponent plays perfectly and making no attempt to exploit any possible weakness. This nicely bounds our results in these plots. Results are given for Best Response, BBR, MAP, Thompson's, Opti, and Frequentist.

As we would expect, the Bayesian players do well against opponents drawn from their prior, with little difference between the three response types in terms of bankroll. The winning rates show that MAP and Thompson's converge within the first ten hands, whereas BBR is more erratic and takes longer to converge. The uninformed Frequentist is clearly behind. The independent Dirichlet prior is very broad, admitting a wide variety of opponents. It is encouraging that the Bayesian approach is able to exploit even this weak information to achieve a better result. However, it is unfair to make strong judgements on the basis of these results since, in general, playing versus its prior is the best possible scenario for the Bayesian approach.

Figures 4 and 5 show bankroll and winning rate results for BBR, MAP, Thompson's, Opti, and Frequentist versus Opti on Leduc hold'em. Note that, on average, a positive bankroll again Opti is impossible, although sample variance allows for it in our experiments. From these plots we can see that the three Bayesian approaches behave very similarly. This is due to the fact that the posterior distribution over our sample of strategies concentrates very rapidly on a single strategy. Within less than 20 hands, one strategy dominates the rest. This means that the three responses become very similar (Thompson's is almost certain to pick the MAP strategy, and BBR puts most of its weight on the MAP strategy). Larger sample sizes would mitigate this effect. The winning rate graphs also show little difference between the three Bayesian players.

Frequentist performs slightly worse than the Bayes approaches. The key problem with it is that it can form models of the opponent that are not consistent with any behavioral strategy (*e.g.*, it can be led to believe that its opponent can always show a winning hand). Such incorrect beliefs, untempered by any prior, can lead it to fold with high probability in certain situations. Once it starts folding, it can never make the observations required to correct its mistaken belief. Opti, of course, breaks even against itself. On the whole, independent Dirichlet distributions are a poor prior for the Opti solution, but we see a slight improvement over the pure frequentist approach.

Our final Leduc results are shown in Figure 6, playing against the Frequentist opponent. These results are included for the sake of interest. Because the Frequentist opponent is not stationary, it violates the assumptions upon which the Bayesian (and, indeed, the Frequentist) player are based. We cannot drawn any real conclusions from this data. It is interesting, however, that the BBR response is substantially worse than MAP or Thompson's.

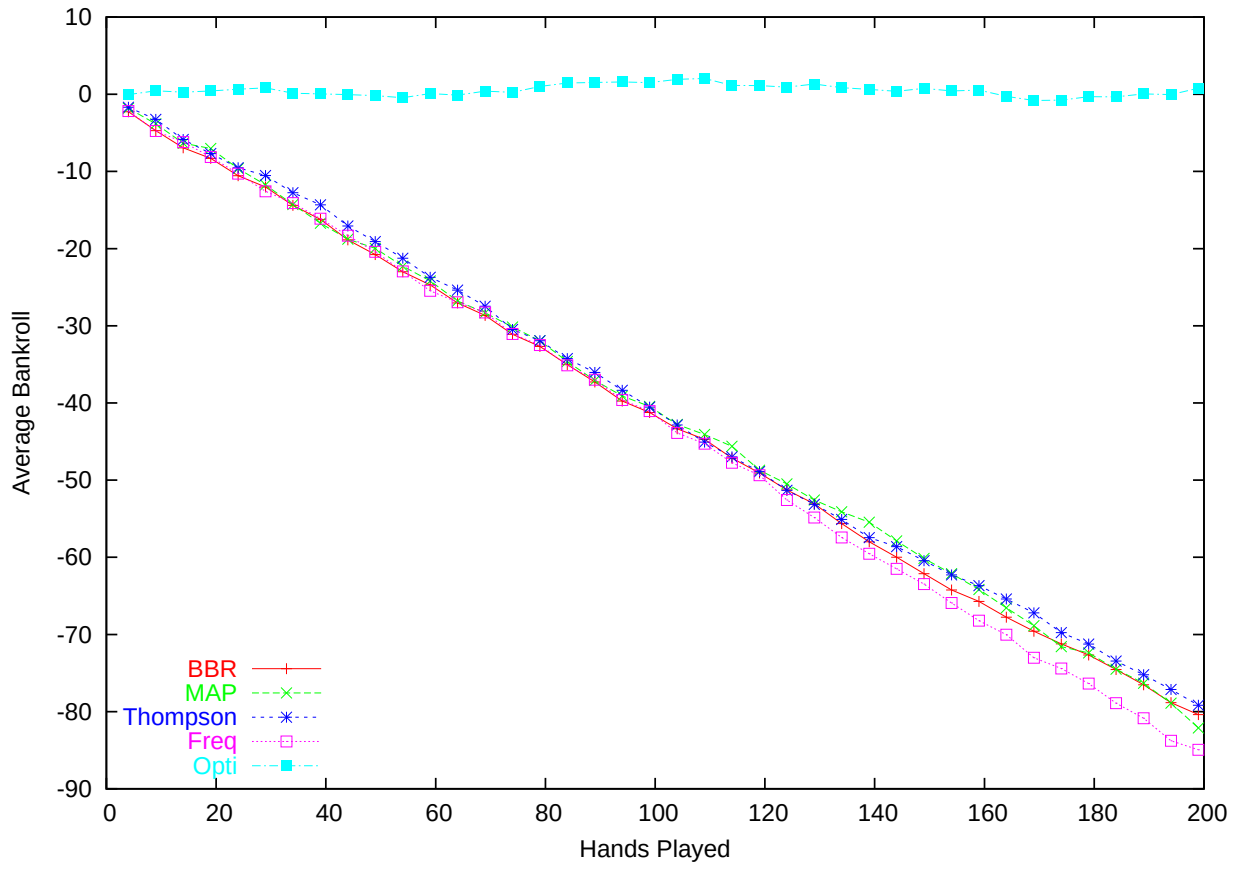

Figure 4: Leduc hold'em: Avg. Bankroll per hands played for BBR, MAP, Thompson's, Opti, and Frequentist vs. Opti.

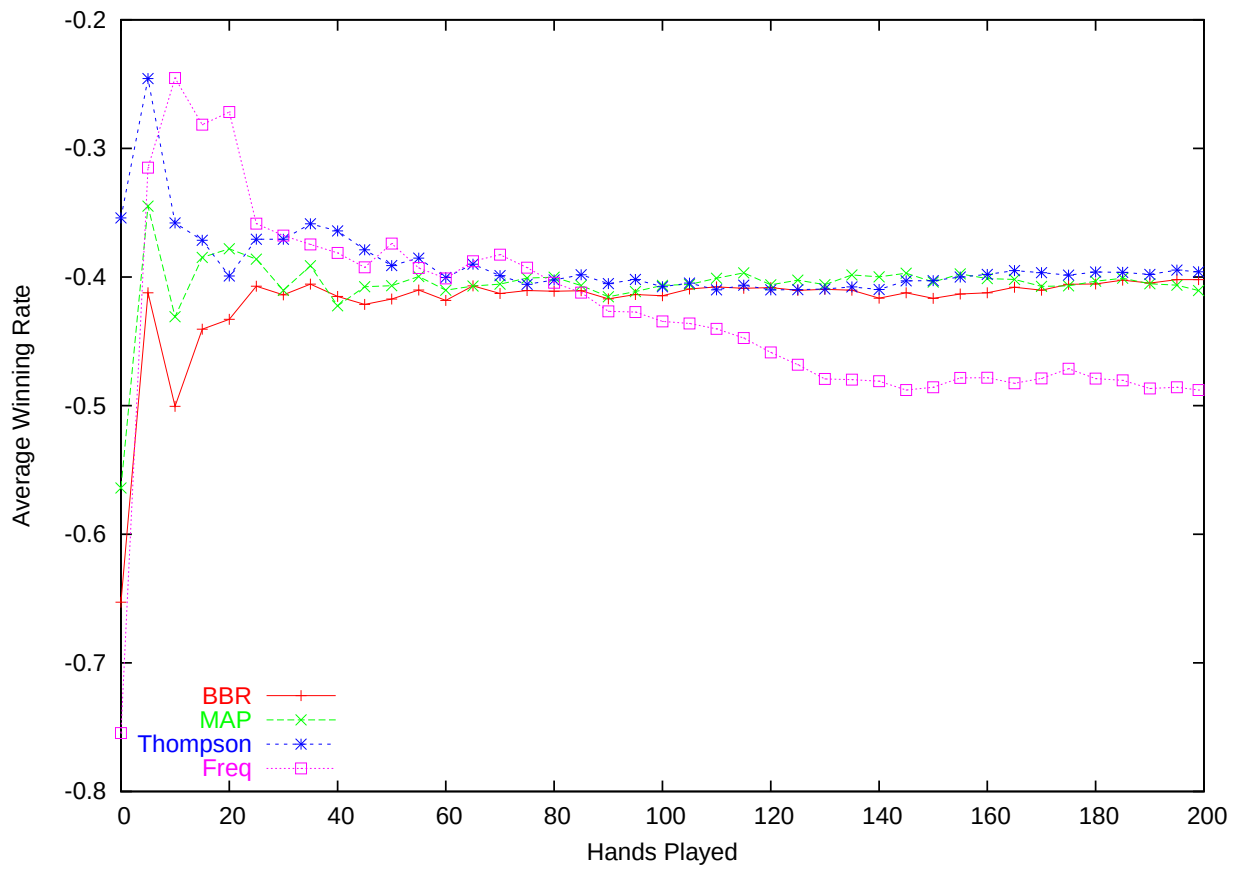

Figure 5: Leduc hold'em: Avg. Winning Rate per hands played for BBR, MAP, Thompson's, Opti, and Frequentist vs. Opti.

It seems likely that the posterior distribution does not converge quickly against a non-stationary opponent, leading BBR to respond to several differing strategies simulataneously. Because the prior is independent for every information set, these various strategies could be giving radically different advice in many contexts, preventing BBR from generating a focused response. MAP and Thompson's necessarily generate more focused responses. We show winning rates in Figure 7 for the sake of completeness, with the same caveat regarding non-stationarity.

### 7.2 Texas Hold'em

Figure 8 show bankroll results for Thompson's, Opti, and Frequentist versus opponents sampled from the informed prior for Texas hold'em. Here Thompson's and Frequentist give very similar performance, although there is a small advantage to Thompson's late in the run. It is possible that even with the more informed prior, two hundred hands does not provide enough information to effectively concentrate the posterior on good models of the opponent in this larger game. It may be that priors encoding strong correlations between many information sets are required to gain a substantial advantage over the Frequentist approach.

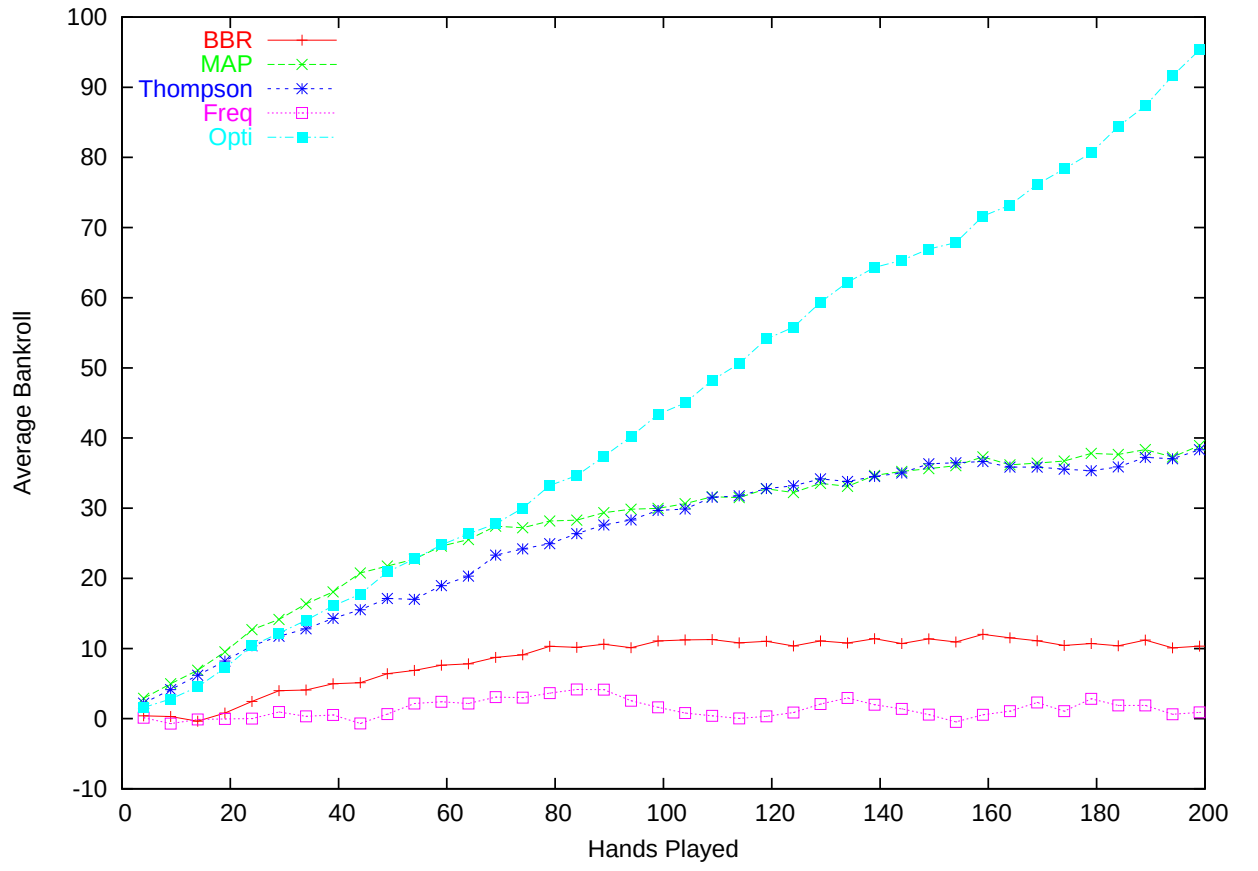

Figure 6: Leduc hold'em: Avg. Bankroll per hands played for BBR, MAP, Thompson's, and Opti vs. Frequentist.

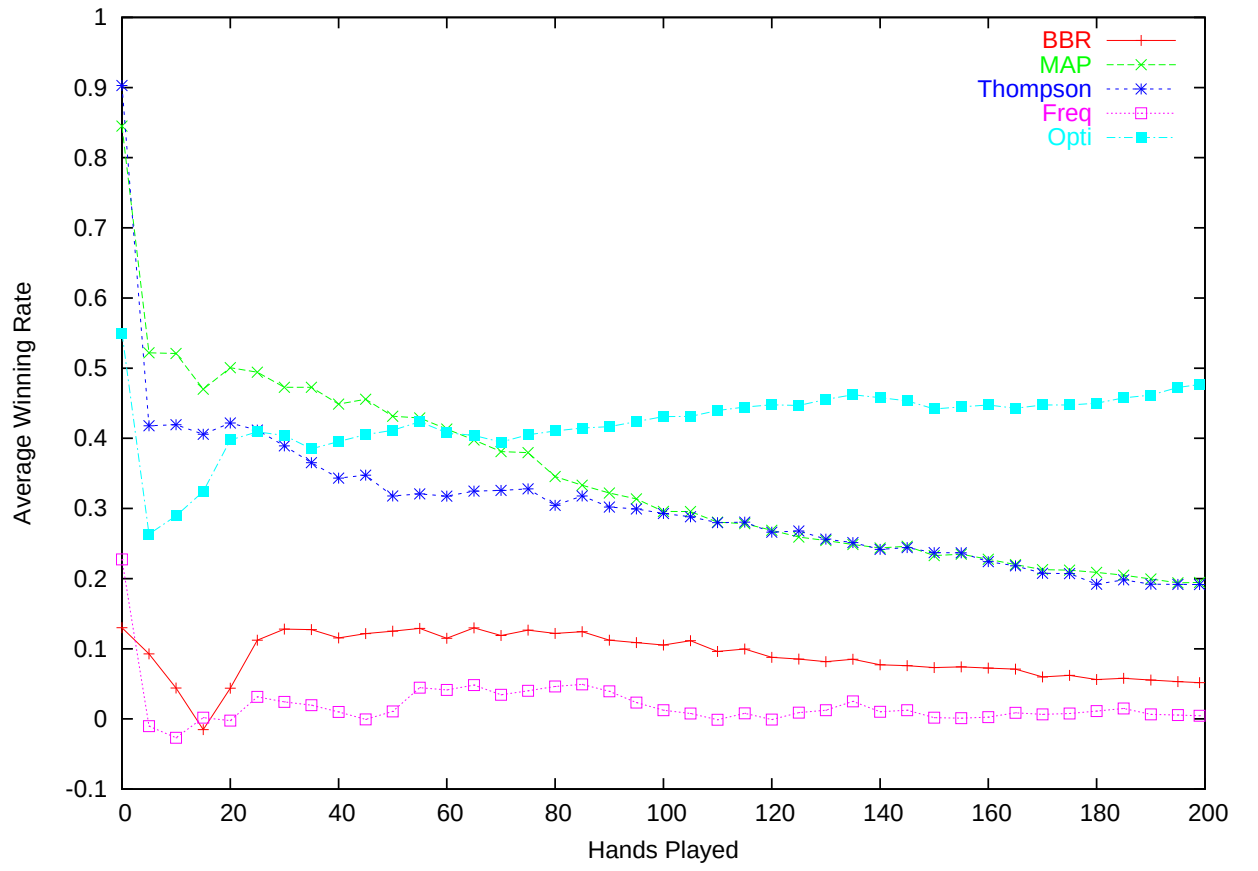

Figure 7: Leduc hold'em: Avg. Winning Rate per hands played for BBR, MAP, Thompson's, and Opti vs. Frequentist.

## 8 Conclusion

This research has presented a Bayesian model for hold'em style poker, fully modelling both game dynamics and opponent strategies. The posterior distribution has been described and several approaches for computing appropriate responses considered. Opponents in both Texas hold'em and Leduc hold'em have been played against using Thompson's sampling for Texas hold'em, and approximate Bayesian best response, MAP, and Thompson's for

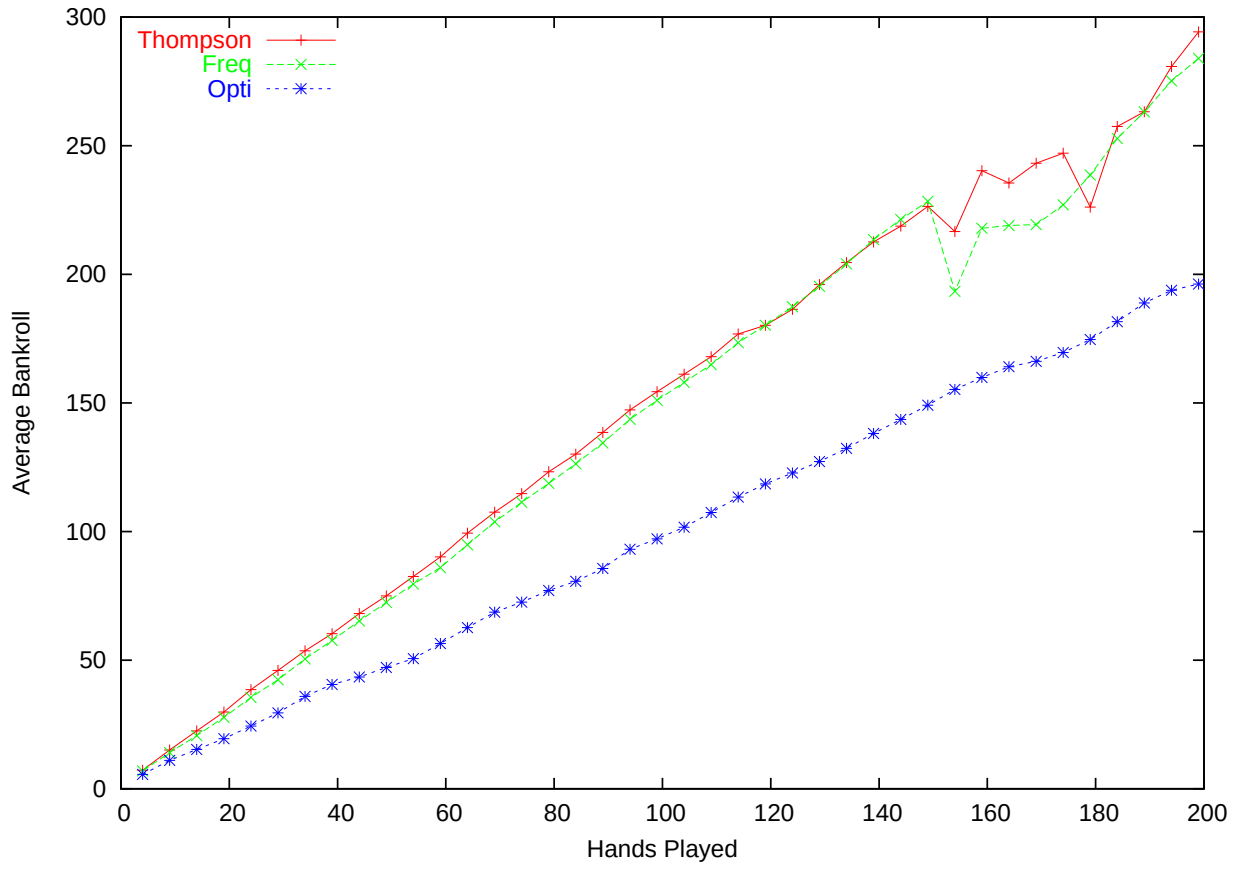

Figure 8: Texas hold'em: Avg. Bankroll per hands played for Thompson's, Frequentist, and Opti vs. Priors.

Leduc hold'em. These results show that, for opponents drawn from our prior, the posterior captures them rapidly and the subsequent response is able to exploit the opponent, even in just 200 hands. On Leduc, the approach performs favourably compared with state-of-the-art opponent modelling techniques against prior-drawn opponents and a Nash equilibrium. Both approaches can play quickly enough for real-time play against humans.

The next major step in advancing the play of these systems is constructing better informed priors capable of modelling more challenging opponents. Potential sources for such priors include approximate game theoretic strategies, data mined from logged human poker play, and more sophisticated modelling by experts. In particular, priors that are capable of capturing correlations between related information sets would allow for generalization of observations over unobserved portions of the game. Finally, extending the approach to non-stationary approaches is under active investigation.

### Acknowledgements

We would like to thank Rob Holte, Dale Schuurmanns, Nolan Bard, and the University of Alberta poker group for their insights. This work was funded by the Alberta Ingenuity Centre for Machine Learning, iCore, and NSERC.